\documentclass[aip,pof,longbibliography,twocolumn,reprint]{revtex4-1} 
\usepackage[cp850]{inputenc}
\usepackage{natbib}
\usepackage{amssymb,amsmath}
\usepackage{epsfig}
\usepackage{bm}
\usepackage{color}

\usepackage{graphicx}

\usepackage{epstopdf}

\usepackage{color}
\newcommand{\vicente}[1]{{ #1}}

\newcommand\beq{\begin{equation}}
\newcommand\eeq{\end{equation}}
\newcommand\beqa{\begin{eqnarray}}
\newcommand\eeqa{\end{eqnarray}}

\newcommand{\al}{\alpha}

\begin{document}
\title{Assessment of kinetic theories for moderately dense granular binary mixtures: Shear viscosity coefficient}
\author{Mois\'es Garc\'ia Chamorro}
\email{moises@unex.es}
\affiliation{Departamento de F\'{\i}sica,
Universidad de Extremadura, Avda. de Elvas s/n, E-06006 Badajoz, Spain}
\author{Vicente Garz\'{o}}
\email{vicenteg@unex.es} \homepage{http://www.unex.es/eweb/fisteor/vicente/}
\affiliation{Departamento de F\'{\i}sica and Instituto de Computaci\'on Cient\'{\i}fica Avanzada (ICCAEx), Avda. de Elvas s/n,, Universidad de Extremadura, E-06006 Badajoz, Spain}

\begin{abstract}

Two different kinetic theories [J. Solsvik and E. Manger \vicente{(SM)}, Phys. Fluids \textbf{33}, 043321 (2021) and V. Garz\'o, J. W. Dufty, and C. M. Hrenya \vicente{(GDH)}, Phys. Rev. E \textbf{76}, 031303 (2007)] are considered to determine the shear viscosity $\eta$ for a moderately dense granular binary mixture of smooth hard spheres. The mixture is subjected to a simple shear flow and heated by the action of an external driving force (Gaussian thermostat) that exactly compensates the energy dissipated in collisions. The set of Enskog kinetic equations is the starting point to obtain the dependence of $\eta$ on the control parameters of the mixture: solid fraction, concentration, mass and diameter ratios, and coefficients of normal restitution. While the expression of $\eta$ found in the SM-theory is based on the assumption of Maxwellian distributions for the velocity distribution functions of each species, the GDH-theory solves the Enskog equation by means of the Chapman--Enskog method to first order in the shear rate. To assess the accuracy of both kinetic theories, the Enskog equation is numerically solved by means of the direct simulation Monte Carlo (DSMC) method. The simulation is carried out for a mixture under simple shear flow, using the thermostat to control the cooling effects. Given that the SM-theory predicts a vanishing kinetic contribution to the shear viscosity, the comparison between theory and simulations is essentially made at the level of the collisional contribution $\eta_c$ to the shear viscosity. The results clearly show that the GDH-theory compares with simulations much better than the SM-theory over a wide range of values of the coefficients of restitution, the volume fraction, and the parameters of the mixture (masses, diameters, and concentration).

\end{abstract}

\draft

\date{\today}

\maketitle

\section{Introduction}
\label{sec1}

The determination of the transport coefficients of polydisperse granular mixtures (namely, mixtures constituted by smooth inelastic hard spheres of different masses, diameters, and coefficients of restitution) is still a challenging objective. There are likely two main reasons for which the above target is quite complex. First, there is a large number of parameters and transport coefficients involved in the description of granular mixtures. Second, there is a wide array of intricacies and uncontrolled approximations arising in the derivation of the corresponding kinetic theories.

Therefore, due to the above difficulties, many of the previous attempts for obtaining the Navier--Stokes transport coefficients of granular mixtures \cite{JM89,AW98,WA99,SGNT06} consider mixtures constituted by nearly elastic spheres. In this limit case it is justified to assume the equipartition of the total granular kinetic energy in the homogeneous cooling
state (HCS). This means that the zeroth-order contributions $T_i^{(0)}$ to the partial temperatures $T_i$ of each species are equal to the (global) granular temperature $T$.

However, as theoretical calculations, \cite{MP99,GD99b} computer simulations, \cite{MG02,BT02,BT02b,DHGD02,PMP02,KT03,WJM03,BRM05,SUKSS06} and real experiments \cite{WP02,FM02} have shown, the assumption of energy equipartition between mechanically different particles only occurs when the collisions are perfectly elastic. A general conclusion of the above works is that the departure of energy equipartition increases as inelasticity increases and the mechanical differences between the particles of each species become more significant (specially when the masses are more disparate).

Although the breakdown of energy equipartition in granular mixtures was pointed out independently by Jenkins and Mancini \cite{JM87} and Zamankhan \cite{Z95} (this author noted energy nonequipartition but assumed equal partial temperatures for studying rheology in sheared granular mixtures), to the best of our knowledge, the impact of energy nonequipartition on transport properties in granular mixtures was analyzed for the first time by Huilin \emph{et al.} \cite{HWRLG00,HGM01} They proposed a two-temperature kinetic theory where the one-particle velocity distribution function of each species $f_i(\mathbf{r}, \mathbf{v};t)$ is a Maxwellian distribution at the partial temperature $T_i^{(0)}$, even for \emph{inhomogeneous} states. Although this approximation could provide acceptable estimates of the collisional transfer contributions to the fluxes and the cooling rate, it predicts vanishing Navier-Stokes transport coefficients in the low-density limit. This is an important drawback of these theories.\cite{HZXKW22} Based on the Maxwellian approximation for $f_i$, Solsvik and Manger \vicente{(SM)} \cite{SM21a,SM21} have recently proposed a kinetic theory \vicente{(hereafter referred to as the SM-theory)} where the distributions $f_i$ take into account not only the temperature differences of the species but also the differences in the mean flow velocities $\mathbf{U}_i$ of the species. Within this approach, the authors \cite{SM21} obtain corrections to the collisional contributions to the momentum and heat fluxes, which are of the order $|\mathbf{U}_i-\mathbf{U}_j|^2$ and $|\mathbf{U}_i-\mathbf{U}_j|^4$.

A different approach for determining the Navier--Stokes transport coefficients for moderately dense granular mixtures have been developed by Garz\'o, Dufty, and Hrenya \vicente{(GDH)}. \cite{GDH07,GHD07} These authors solve the Enskog kinetic equation by means of the Chapman--Enskog method \cite{CC70} adapted to dissipative dynamics \vicente{(hereafter the theory proposed by GDH will be referred to as the GDH-theory)}. In the first-order of spatial gradients, as for molecular mixtures, \cite{LCK83} the transport coefficients are defined in terms of the solutions of a set of coupled linear integral equations. These equations are approximately solved by considering the leading terms in a Sonine polynomial expansion of the first-order distribution functions. Thus, explicit expressions for the transport coefficients and the cooling rate are obtained in terms of the parameter space of the mixture (masses and diameters, concentrations, solid volume fraction, and coefficients of restitution). These expressions apply in principle to arbitrary values of the coefficients of restitution and are not limited to specific values of the remaining parameters of the mixture. In fact, the GDH-theory reduces in the limit of mechanically equivalent particles to well established kinetic theory models \cite{GD99a,L05} for monocomponent granular gases. In addition, the GDH-theory compares in general very well with computer simulation results obtained for the tracer diffusion coefficient \cite{GM04} and the shear viscosity coefficient of a heated granular binary mixture. \cite{GM07}

On the other hand, given that the SM-theory \cite{SM21} can be only reliable for obtaining the collisional contributions to the transport coefficients, an interesting problem is to assess the degree of accuracy of the SM and GDH theories by comparing their predictions (for the collisional coefficients) against computer simulations. Although the predictions of the GDH-theory for the shear viscosity coefficient $\eta$ were already tested with simulations in Ref.\ \onlinecite{GM03}, only simulation data for the kinetic $\eta_k$ and global shear viscosity $\eta$ were reported in this paper. Thus, it could be convenient to perform new simulations where the dependence of the collisional shear viscosity coefficient $\eta_c$ on the parameter space of the mixture were widely analyzed. This would allow us to asses the degree of accuracy of the SM and GDH theories for dense granular mixtures. The objective of this paper is to carry out new simulations for determining $\eta_c$ and compare them with those predicted by the SM and GDH theories. This will allow us to gauge the strengths and weaknesses of both kinetic theories.

As in the simulations performed in Ref.\ \onlinecite{GM03}, we consider here a particular hydrodynamic state: the so-called simple (or uniform) shear flow (SSF) state. This state is characterized by constant partial densities $n_i$, uniform granular temperature $T$, and a linear velocity profile $U_{1,\lambda}=U_{2,\lambda}=a_{\lambda \beta} r_\beta$, where $a_{\lambda \beta}=a\delta_{\lambda x}\delta_{\beta y}$, $a$ being the \emph{constant} shear rate. In the case of a molecular mixture (elastic collisions), unless a thermostating mechanism is introduced, the temperature grows in time due to the viscous heating term $-a P_{xy}$ ($P_{xy}<0$ is the $xy$-component of the pressure tensor). A consequence of the viscous heating effect is that the effective collision frequency for hard spheres $\nu(t)$ (which is proportional to $\sqrt{T(t)}$) increases with time and so, the \emph{reduced} shear rate $a^*(t)=a/\nu(t)$ tends to zero for times longer than the (effective) mean free time $\nu^{-1}$. Thus, for sufficiently long times, the system achieves a regime described by linear hydrodynamics and the Navier--Stokes shear viscosity coefficient $\eta$ can be measured in computer simulations. This procedure was followed many years ago by Naitoh and Ono \cite{NO79} for getting $\eta$ for molecular hard-spheres gases. In the case of granular gases, unfortunately the relation between the temperature and the shear viscosity is not as simple as for molecular gases due to the presence of the collisional term arising from  inelasticity in collisions. However, if there is a thermostat that injects energy to the system that compensates for the collisional energy loss, then the viscous heating term heats the system (as for molecular gases) and one can identify the shear viscosity in the limit $a^*\to 0$. Here, as in Ref.\ \onlinecite{GM03}, we consider the Gaussian thermostat (external force proportional to the particle velocity). In the absence of a shear field, this thermostat (which is usually employed in nonequilibrium molecular dynamics simulations \cite{EM90}) has the advantage that it plays a neutral role in the dynamics of the system. \cite{MS00}

The plan of the paper is as follows. In Sec.\ \ref{sec2}, the Enskog kinetic equation in the SSF state is introduced. Expressions of the pressure tensor and the cooling rate in the local Lagrangian frame where the SSF is \emph{homogeneous} are also displayed. Sections \ref{sec3} and \ref{sec4} provide the results obtained for the shear viscosity in the (driven) SSF from the SM and GDH theories, respectively. Section \ref{sec5} deals with the application of the direct simulation Monte Carlo (DSMC) method \cite{B94} (the extension of this method to dense gases is usually referred to as the ESMC method) to the SSF with thermostat. The theoretical results obtained from the SM and GDH theories for the collisional shear viscosity coefficient $\eta_c$ are compared in Sec.\ \ref{sec6} with computer simulations. The results show that the GDH-theory compares with simulations much better than the SM-theory. We close the paper in Sec.\ \ref{sec7} with some concluding remarks.


\section{Enskog kinetic theory. Simple shear flow state}
\label{sec2}

\subsection{Enskog equation for granular mixtures}

We consider a granular binary mixture of inelastic hard disks ($d=2$) or spheres ($d=3$) of masses $m_{1}$ and $m_{2} $, and diameters $\sigma _{1}$ and $\sigma _{2}$. We assume that the spheres are completely smooth so that the inelasticity of collisions among all pairs is characterized by three independent constant (positive) coefficients of normal restitution $\alpha_{11}$, $\alpha_{22}$, and $ \alpha_{12}=\alpha_{21}$. Here, $\alpha_{ij}\leq 1$ is the restitution coefficient for collisions between particles of species $i$ and $j$. The case $\al_{ij}=1$ corresponds to elastic collisions (molecular mixtures of hard spheres).

Due to the inelastic character of collisions, it is quite usual in experiments to supply energy to the system to balance the collisional loss of energy. This can be done by driving the system through the boundaries \cite{YHCMW02} or alternatively by bulk driving, as in air-fluidized beds. \cite{SGS05,AD06} However, these ways of supplying energy produce in many cases strong spatial gradients in the bulk domain and so, the Navier--Stokes description fails. For this reason, it is frequent in computer simulations \cite{PLMPV98,SVGP10,GSVP11,FAZ09,SSW13,KG14,ChVG15,GGG21} to heat the system homogenously by the action of an external driving force. Borrowing a terminology used in nonequilibrium molecular-dynamics simulations of
ordinary (or molecular) fluids, \cite{EM90} these types of external forces are called \emph{thermostats}. In the present paper, for simplicity, we introduce the so-called Gaussian thermostat, namely, a deterministic external force proportional to the peculiar velocity $\mathbf{V}$. This sort of thermostat has been frequently employed in nonequilibrium molecular dynamics simulations of elastic particles. \cite{EM90}

Under the above conditions, the Enskog kinetic equation for the one-particle velocity distribution function of species $i$ ($i=1,2$) is given by
\begin{equation}
\left( \partial _{t}+\mathbf{v}\cdot \nabla \right)
f_{i}+\frac{1}{2}\xi
\frac{\partial}{\partial \mathbf{v}}\cdot \left(\mathbf{V}f_i\right)
=\sum_{j=1}^2J_{ij}^{\text{E}}\left[ {\bf r}, {\bf v}|f_{i}(t),f_{j}(t)\right] \;,
\label{2.1}
\end{equation}
where the constant $\xi$ is chosen to be the same for both species. Here, $\mathbf{V}=\mathbf{v}-\mathbf{U}$,
\beq
\label{2.1.1}
\mathbf{U}=\sum_{i=1}^2 \frac{\rho_i}{\rho}\mathbf{U}_i=\rho^{-1}\sum_{i=1}^2 \int d\mathbf{v}\; m_i \mathbf{v} f_i(\mathbf{v})
\eeq
is the mean flow velocity of the mixture, $\rho=\rho_1+\rho_2$ is the total mass density, $\rho_i=m_i n_i$, and
\beq
\label{2.1.2}
n_i=\int d\mathbf{v}\; f_i(\mathbf{v})
\eeq
is the number density of species $i$. The second equality in Eq.\ \eqref{2.1.1} defines the mean flow velocities $\mathbf{U}_i$ of species $i$. Apart from $n_i$ and $\mathbf{U}$, the other relevant hydrodynamic field is the granular temperature $T$. It is defined as
\beq
\label{2.1.3}
T=\frac{1}{d n} \sum_{i=1}^2 \int d\mathbf{v}\; m_i V^2 f_i(\mathbf{v}),
\eeq
where $n=n_1+n_2$ is the total number density.

In Eq.\ \eqref{2.1}, the Enskog collision operator $J_{ij}^{\text{E}}[f_i,f_j]$ is \cite{G19}
\begin{eqnarray}
\label{2.2}
& & J_{ij}^{\text{E}}\left[ {\bf r}, {\bf v}_{1}|f_{i},f_{j}\right] =\sigma _{ij}^{d-1}\int d\mathbf{v}
_{2}\int d\widehat{\boldsymbol {\sigma}}\,\Theta (\widehat{\boldsymbol {\sigma}}
\cdot {\bf g})(\widehat{\boldsymbol {\sigma}}\cdot \mathbf{g})  \nonumber
\\ &&\times \left[ \alpha_{ij}^{-2}\chi_{ij}({\bf r},{\bf r}-{\boldsymbol{\sigma}}_{ij})
f_i(\mathbf{r}, \mathbf{v}_1'';t)f_j(\mathbf{r}-{\boldsymbol {\sigma}}_{ij}, \mathbf{v}_2'';t)\right.
\nonumber\\
& & \left.
-\chi_{ij}(\mathbf{r},\mathbf{r}+{\boldsymbol {\sigma}}_{ij})
f_i(\mathbf{r}, \mathbf{v}_1;t)f_j(\mathbf{r}+{\boldsymbol {\sigma}}_{ij}, \mathbf{v}_2;t)\right].
\end{eqnarray}
Here, ${\boldsymbol {\sigma}}_{ij}=\sigma_{ij} \widehat{\boldsymbol {\sigma}}$ with $\sigma _{ij}=\left( \sigma_{i}+\sigma _{j}\right) /2$ and $\widehat{\boldsymbol {\sigma}}$ is a unit vector directed along the line of centers from the sphere of species $i$ to the sphere of species $j$ upon collision (i.e. at contact). In addition,  $\Theta $ is
the Heaviside step function, and $\mathbf{g}=\mathbf{v}_{1}-\mathbf{v}_{2}$ is the relative velocity of the colliding pair. The double
primes on the velocities denote the initial values $\{\mathbf{v}_{1}'',
\mathbf{v}_{2}''\}$ that lead to $\{\mathbf{v}_{1},\mathbf{v}_{2}\}$
following a binary collision:
\begin{equation}
\label{2.3.0}
\mathbf{v}_{1}''=\mathbf{v}_{1}-\mu_{ji}\left(1+\alpha_{ij}^{-1}\right)
(\widehat{{\boldsymbol {\sigma}}}\cdot \mathbf{g})\widehat{{\boldsymbol {\sigma}}},
\eeq
\beq
\mathbf{v}_{2}''=\mathbf{v}_{2}+\mu_{ij}\left( 1+\alpha
_{ij}^{-1}\right) (\widehat{{\boldsymbol {\sigma}}}\cdot \mathbf{g})\widehat{
{\boldsymbol {\sigma}}}, \label{2.3}
\end{equation}
where $\mu _{ij}=m_{i}/\left( m_{i}+m_{j}\right)$. Inversion of the collision rules \eqref{2.3.0} and \eqref{2.3} provides
the form of the so-called \emph{direct} collisions, namely, collisions where the pre-collisional velocities $(\mathbf{v}_1, \mathbf{v}_2)$ lead to the post-collisional velocities $(\mathbf{v}_1', \mathbf{v}_2')$:
\begin{equation}
\label{2.3.1}
\mathbf{v}_{1}'=\mathbf{v}_{1}-\mu_{ji}\left(1+\alpha_{ij}\right)
(\widehat{{\boldsymbol {\sigma}}}\cdot \mathbf{g})\widehat{{\boldsymbol {\sigma}}},
\eeq
\beq
\mathbf{v}_{2}'=\mathbf{v}_{2}+\mu_{ij}\left( 1+\alpha
_{ij}\right) (\widehat{{\boldsymbol {\sigma}}}\cdot \mathbf{g})\widehat{
{\boldsymbol {\sigma}}}. \label{2.3.2}
\end{equation}

The quantity $\chi_{ij}[{\bf r},{\bf r}+{\boldsymbol {\sigma}}_{ij}|\{n_\ell\}] $ is the equilibrium pair correlation function of two hard spheres, one of species $i$ and the other of species $j$, at contact, i.e., when the distance between their centers is $\sigma_{ij}$. In the original phenomenological kinetic theory of Enskog\cite{FK72} (which is usually referred to as the standard Enskog theory), the $\chi_{ij}$ are the same {\em functions} of the densities $\{n_\ell\}$ as in a fluid mixture in {\em uniform} equilibrium. On the other hand, this choice for $\chi_{ij}$ leads to some inconsistencies with irreversible thermodynamics.  In order to fix this conceptual problem, van Beijeren and Ernst \cite{BE73c} proposed  an alternative generalization to the Enskog equation for mixtures, which is usually referred to as the revised Enskog theory (RET).  In the RET, the $\chi_{ij}$ are the same {\em functionals} of the densities $\{n_\ell\}$ as in a fluid in {\em nonuniform} equilibrium. This fact increases considerably the technical difficulties involved in the derivation of the general hydrodynamic equations from the RET, \cite{LCK83} unless the partial densities are uniform as occurs in the SSF state.

\subsection{Simple shear flow}

As mentioned in section \ref{sec1}, we want to solve the Enskog equation \eqref{2.1} in the SSF state. At a macroscopic level, the SSF is characterized by uniform partial densities $n_i$ and temperature $T$ and a linear velocity profile given by
\beq
\label{2.4}
\mathbf{U}_1=\mathbf{U}_2=\mathbf{U}=\mathsf{a}\cdot \mathbf{r}, \quad a_{\lambda \beta}=a\delta_{\lambda x} \delta_{\beta y},
\eeq
where $a$ is the \emph{constant} shear rate. In the SSF, the mass and heat fluxes vanish for symmetry reasons and the only flux of the problem is the (uniform) pressure tensor $\mathsf{P}$. For moderate densities, $\mathsf{P}$ has kinetic and collisional contributions. The only relevant hydrodynamic balance equation is that for the temperature $T(t)$. This equation can be deduced by multiplying both sides of Eq.\ \eqref{2.1} by $\frac{1}{2}m_i v^2$, integrating over $\mathbf{v}$, and summing over $i$. It is given by
\beq
\label{2.5}
\partial_t T+\frac{2}{d n} a P_{xy}=-\left(\zeta-\xi\right) T,
\eeq
where $\zeta$ is the cooling rate. This quantity provides the rate of kinetic energy dissipated by inelastic collisions. The expressions of $\mathsf{P}$ and $\zeta$ in the SSF will be displayed below.

It is worthwhile noting that if one chose $\xi=\zeta$ in Eq.\ \eqref{2.5}, then this macroscopic balance equation looks like the energy equation in the SSF state for molecular mixtures. However, in the limit $a^*\to 0$, the corresponding expression of the shear viscosity coefficient differs from the one obtained for a mixture of elastic collisions.

At a microscopic level, the SSF becomes a homogeneous state in the local Lagrangian frame defined by the variables $\mathbf{V}=\mathbf{v}-\mathsf{a}\cdot \mathbf{r}$ and $\mathbf{R}=\mathbf{r}-\mathsf{a}\cdot \mathbf{r} t$. \cite{DSBR86} In this frame,the velocity distribution functions are uniform [$f_i({\bf r},{\bf v},t)=f_i({\bf V},t)$] and the Enskog equation reads
\begin{equation}
\label{2.6}
\partial_{t}f_i-aV_y\frac{\partial}{\partial V_x}
f_{i}+\frac{1}{2}\xi \frac{\partial}{\partial {\bf V}}\cdot \left({\bf V}f_i\right)
=\sum_{j=1}^2J_{ij}^{\text{E}}\left[ {\bf V}|f_{i}(t),f_{j}(t)\right] \;,
\end{equation}
where the operator $J_{ij}^{\text{E}}\left[ {\bf V}|f_{i}(t),f_{j}(t)\right] $ becomes \cite{GM03}
\begin{eqnarray}
\label{2.7}
J_{ij}^{\text{E}}\left[ {\bf V}_{1}|f_{i},f_{j}\right] &=&\sigma _{ij}^{d-1}\chi_{ij}\int d{\bf V}
_{2}\int d\widehat{\boldsymbol {\sigma}}\,\Theta (\widehat{\boldsymbol {\sigma}}
\cdot {\bf g})(\widehat{\boldsymbol {\sigma}}\cdot {\bf g})  \nonumber \\ &&\times \left[ \alpha _{ij}^{-2} f_i({\bf V}_1',t)f_j({\bf V}_2'+a\sigma_{ij}\widehat{\sigma}_y{\widehat{{\bf x}}} ,t)\right.\nonumber\\
& & \left.
-f_i( {\bf V}_1,t)f_j({\bf V}_2-a\sigma_{ij}\widehat{\sigma}_y{\widehat{{\bf x}}},t)\right].
\end{eqnarray}
Note that the functions $\chi_{ij}$ are uniform in the SSF problem. As said before, the pressure tensor has kinetic and collisional transfer contributions:
\beq
\label{2.8}
\mathsf{P}=\mathsf{P}^{\text{k}}+\mathsf{P}^{\text{c}}.
\eeq
The kinetic contribution $\mathsf{P}^{\text{k}}$ is
\beq
\label{2.9}
\mathsf{P}^{\text{k}}=\sum_{i=1}^2\,\int d\mathbf{V}\, m_i \mathbf{V}\mathbf{V} f_i(\mathbf{V}),
\eeq
while the collisional transfer contribution $\mathsf{P}^\text{c}$ in the Lagrangian frame is given by \cite{GM03}
\begin{eqnarray}
\label{2.10}
\mathsf{P}^{\text{c}}&=&\sum_{i=1}^2\sum_{j=1}^2 m_{ij}\chi_{ij}\sigma_{ij}^d
\frac{1+\alpha_{ij}}{2}\int d{\bf V}_1\int d{\bf V}_2\int d\widehat{\boldsymbol {\sigma}}
\nonumber\\
& &
\times \Theta (\widehat{\boldsymbol {\sigma}} \cdot {\bf g})(\widehat{\boldsymbol {\sigma}}\cdot {\bf g})^2
\widehat{\boldsymbol {\sigma}}\widehat{\boldsymbol {\sigma}}
f_i\left({\bf V}_1+a\sigma_{ij}\widehat{\sigma}_y{\widehat{{\bf x}}},t\right)f_j({\bf V}_2,t),
\nonumber\\
\end{eqnarray}
where $m_{ij}=m_im_j/(m_i+m_j)$. The cooling rate $\zeta$ is \cite{GM03}
\begin{eqnarray}
\label{2.11}
\zeta&=&\frac{1}{2dnT}\sum_{i=1}^2\sum_{j=1}^2m_{ij}\chi_{ij}\sigma_{ij}^{d-1}
(1-\alpha_{ij}^2)\nonumber\\
& & \times \int d{\bf V}_1\int d{\bf V}_2\int d\widehat{\boldsymbol {\sigma}}\,\Theta (\widehat{\boldsymbol {\sigma}}
\cdot {\bf g})(\widehat{\boldsymbol {\sigma}}\cdot {\bf g})^3  \nonumber\\
& & \times
f_i\left({\bf V}_1+a\sigma_{ij}\widehat{\sigma}_y{\widehat{{\bf x}}},t\right)f_j({\bf V}_2,t).
\end{eqnarray}

Equations \eqref{2.8}--\eqref{2.11} provide the expressions of the pressure tensor and the cooling rate in terms of the velocity distribution functions $f_i(\mathbf{V};t)$ in the SSF state. Needless to say, it still remains to determine $f_i(\mathbf{V};t)$ to compute the corresponding velocity integrals and get the above quantities. Based on symmetry considerations, to first-order in the shear rate, the pressure tensor $\mathsf{P}^{(1)}$ is
\beq
\label{3.1}
P_{\lambda \beta}^{(1)}=-\eta a \left(\delta_{\lambda x} \delta_{\beta y}+\delta_{\lambda y} \delta_{\beta x}\right),
\eeq
where $\eta$ is the shear viscosity coefficient $\eta$. In this paper, we consider two different kinetic theories to determine $\eta$.

\section{SM-Kinetic theory}
\label{sec3}


The SM-theory \cite{SM21} is based on a simple approximation: the distributions $f_i(\mathbf{V};t)$ are assumed to be Maxwellian distributions $f_{i,\text{M}}(\mathbf{V};t)$:
\beq
\label{3.2}
f_{i,\text{M}}(\mathbf{V};t)=n_i\left(\frac{m_i}{2\pi T_i^{(0)}}\right)^{d/2} \exp\left(-\frac{m_i V^2}{2T_i^{(0)}}\right),
\eeq
where $T_i^{(0)}$ is the zeroth-order contribution to the partial temperature of species $i$. Upon writing Eq.\ \eqref{3.2} we have made use of the fact that the velocity differences $|\mathbf{U}_i-\mathbf{U}_j|$ vanish in the SSF. According to the approximation \eqref{3.2}, the kinetic contribution $\mathsf{P}^\text{k}=\mathsf{0}$ in the SM-theory and the kinetic shear viscosity vanishes ($\eta_\text{k}=0$). This is of course a deficiency of the SM-theory which is not able to capture the kinetic transfer contributions to the shear viscosity, which are different from zero even for granular mixtures at low-density. \cite{GD02,GMD06} This means that this theory can be only seen as a valuable approach for estimating the collisional transfer contribution $\eta_\text{c}$ to $\eta$. According to Eqs.\ (41) and (77) of Ref.\ \onlinecite{SM21} and Eq.\ \eqref{3.1}, $\eta_\text{c}$ for hard spheres ($d=3$) can be identified as \cite{SM21,G21}
\beqa
\label{3.3}
\eta_\text{c}^{\text{SM}}&=&\frac{\sqrt{2\pi}}{15}\sum_{i=1}^2\sum_{j=1}^2 n_i n_j \sigma_{ij}^4\chi_{ij} m_{ij}^2 (1+\al_{ij})\nonumber\\
& & \times \Bigg(\frac{T_i^{(0)}}{m_i}+\frac{T_j^{(0)}}{m_j}\Bigg)^{3/2}\Bigg(\frac{1}{T_i^{(0)}}+\frac{1}{T_j^{(0)}}\Bigg).
\eeqa

It is interesting to note that the expression \eqref{3.3} slightly differs from the one obtained by replacing $f_i(\mathbf{V})$ by $f_{i,\text{M}}(\mathbf{V})$ in Eq.\ \eqref{2.10} and performing the corresponding integrals in velocity space. In the linear order of the shear rate, after some algebra, one gets the following expression for the collisional shear viscosity $\eta_\text{c}\simeq \eta_\text{c}^{\text{M}}$:
\beqa
\label{3.4}
\eta_\text{c}^{\text{M}}&=&
\frac{\sqrt{2}\pi^{(d-1)/2}}{d (d+2)\Gamma\left(\frac{d}{2}\right)}\sum_{i=1}^2\sum_{j=1}^2 n_i n_j \sigma_{ij}^{d+1}\chi_{ij} m_{ij} (1+\al_{ij})\nonumber\\
& & \times \Bigg(\frac{T_i^{(0)}}{m_i}+\frac{T_j^{(0)}}{m_j}\Bigg)^{1/2}.
\eeqa
For elastic collisions ($\al_{ij}=1$), $T_1^{(0)}=T_2^{(0)}=T$ and so, Eqs.\ \eqref{3.3} and \eqref{3.4} agree for a three-dimensional ($d=3$) system.

\section{GDH-kinetic theory}
\label{sec4}

In contrast to the SM-theory, the GDH-theory\cite{GDH07} solves the Enskog equation \eqref{2.6} by means of the Chapman--Enskog method. \cite{CC70} Since we want to get the shear viscosity coefficient in the driven case when the collisional cooling is exactly compensated for by the energy supplied to the mixture by the external driving force, then we take $\xi=\zeta$ in Eq.\ \eqref{2.6}. With this choice, according to Eq.\ \eqref{2.5}, the temperature increases in time due to the viscous heating term $-a P_{xy}>0$. The determination of $\eta$ under these conditions was carried out years ago in Ref.\ \onlinecite{GM03} for $d=3$. The extension to $d$-dimensional mixtures follows similar steps as those mode in the above work (see the Appendix B of Ref.\ \onlinecite{GM03} for specific technical details on this calculation). We offer here only some partial results for the determination of $\eta$ in the driven SSF.

The Chapman--Enskog method \cite{CC70} provides the \emph{normal} (or hydrodynamic) solution to the Enskog equation \eqref{2.6} as an expansion in powers of the shear rate $a$:
\begin{equation}
\label{3.5}
f_{i}=f_{i}^{(0)}+f_{i}^{(1)}+\cdots,
\end{equation}
where $f_i^{(k)}$ is of order $k$ in $a$. As usual, the time derivatives $\partial_t$, the Enskog collision operator $J_{ij}^{\text{E}}[f_i,f_j]$, and the pressure tensor $\mathsf{P}$ are also expanded as
\begin{equation}
\label{3.6}
\partial_t=\partial_t^{(0)}+ \partial_t^{(1)}+\cdots,\quad J_{ij}^{\text{E}}=J_{ij}^{(0)}+J_{ij}^{(1)}+\cdots,
\end{equation}
\begin{equation}
\label{3.7.0}
\mathsf{P}=\mathsf{P}^{(0)}+\mathsf{P}^{(1)}+\cdots.
\end{equation}
As $\xi=\zeta$ at any order in the shear rate, then $\partial_t^{(0)}T=0$ and
\beq
\label{3.8}
\partial_t^{(1)}T=-\frac{2}{dn}aP_{xy}^{(0)}.
\eeq

\subsection{Zeroth-order approximation}

To zeroth-order in $a$, the Enskog equation \eqref{2.6} reads \cite{GM03}
\begin{equation}
\label{3.9}
 \frac{1}{2}\zeta^{(0)}\frac{\partial}{\partial \mathbf{V}}\cdot \left(\mathbf{V}f_i^{(0)}\right)
=\sum_{j=1}^2J_{ij}^{(0)}[f_i^{(0)},f_j^{(0)}],
\end{equation}
where $\zeta^{(0)}$ is given by Eq.\ \eqref{2.11} with the replacements $f_i\to f_i^{(0)}$, $f_j\to f_j^{(0)}$, and
\begin{eqnarray}
\label{3.10}
J_{ij}^{(0)}[f_i^{(0)},f_j^{(0)}]&=&\chi_{ij}\sigma_{ij}^{d-1}\int d{\bf V}_2 \int
d\widehat{\boldsymbol {\sigma}}\,\Theta (\widehat{\boldsymbol {\sigma}}
\cdot {\bf g})(\widehat{\boldsymbol {\sigma}}\cdot {\bf g}) \nonumber\\
& & \times \left[\alpha_{ij}^{-2}f_i^{(0)}({\bf V}_1')f_j^{(0)}({\bf V}_2')\right.
\nonumber\\
& &\left.
-f_i^{(0)}({\bf V}_1)f_j^{(0)}({\bf V}_2)\right].
\end{eqnarray}

Equation  \eqref{3.11} turns out to be formally {\em identical} to the one obtained in the HCS (i.e., in the unforced case with $\xi=0$). \cite{GD99b,G19} Thus, when one properly scales the velocities $\mathbf{v}$ with the thermal speed $v_\text{th}\propto \sqrt{T(t)}$, there is an exact equivalence between the results derived in the HCS and those obtained when the mixture is driven by the Gaussian thermostat. This is one of the advantages of using this thermostat. On the other hand, this equivalence fails for \emph{inhomogeneous} situations and the external force does not play a neutral role in the evaluation of the transport properties. \cite{DSBR86}

Since the distributions $f_i^{(0)}(\mathbf{V})$ are isotropic in $\mathbf{V}$, then the pressure tensor is diagonal: $P_{\lambda \beta}^{(0)}=p \delta_{\lambda \beta}$, where the hydrostatic pressure $p$ is \cite{G19}
\beq
\label{3.11}
p\sum_{i=1}^2n_iT_i^{(0)}+\frac{\pi^{d/2}}{d\Gamma\left(\frac{d}{2}\right)}
\sum_{i=1}^2\sum_{j=1}^2\sigma_{ij}^d\chi_{ij}n_in_j
\mu_{ji}(1+\alpha_{ij})T_i^{(0)}.
\eeq
Since $\mathsf{P}^{(0)}$ is a diagonal tensor, then $\partial_t^{(1)} T=0$ in accordance with Eq.\ \eqref{3.8}.

Note that the partial temperatures have the constraint
\beq
\label{3.12}
n T=n_1 T_1^{(0)}+n_2 T_2^{(0)}.
\eeq
For elastic collisions ($\al_{ij}=1$), $T_1^{(0)}=T_2^{(0)}=T$ and so, the total kinetic energy is equally distributed between the two species of the mixture. However, for inelastic collisions ($\al_{ij}<1$), the partial temperatures $T_i^{(0)}$ are in general different from the (global) granular temperature $T$ and so, energy equipartition is broken down.

It still remains to get the dependence of the temperature ratio $\gamma\equiv T_1^{(0)}/T_2^{(0)}$ on the parameter space of the mixture. The expression of $\gamma$ will be also used later in both SM-theory and GDH-theory to determine $\eta_c$ in terms of the the parameters of the mixture. The condition for determining the ratio $T_1^{(0)}/T_2^{(0)}$ is \cite{GD99b}
\beq
\label{3.5.0}
\zeta_1^{(0)}=\zeta_2^{(0)}=\zeta^{(0)},
\eeq
where the partial cooling rates $\zeta_i^{(0)}$ are associated to the partial temperatures $T_i^{(0)}$. Here,
\beq
\label{3.6.0}
\zeta^{(0)}=\frac{1}{n T}\sum_{i=1}^2\; n_i T_i^{(0)} \zeta_i^{(0)}.
\eeq
The partial cooling rates $\zeta_i^{(0)}$ are defined as
\beqa
\label{3.6.1}
\zeta_i^{(0)}&=&\frac{1}{2dn_iT_i^{(0)}}\sum_{j=1}^2m_{ij}\chi_{ij}\sigma_{ij}^{d-1}
(1-\alpha_{ij}^2)\nonumber\\
& & \times \int d{\bf V}_1\int d{\bf V}_2\int d\widehat{\boldsymbol {\sigma}}\,\Theta (\widehat{\boldsymbol {\sigma}}
\cdot {\bf g})(\widehat{\boldsymbol {\sigma}}\cdot {\bf g})^3  \nonumber\\
& & \times
f_i^{(0)}\left({\bf V}_1,t\right)f_j^{(0)}({\bf V}_2,t).
\eeqa
A good estimate of $\zeta_i^{(0)}$ can be obtained by considering the Maxwellian approximation \eqref{3.2} for the zeroth-order distributions $f_i^{(0)}({\bf V})$. In this case, the partial cooling rates are \cite{GD99b,G19}
\beqa
\label{3.7}
\zeta_i^{(0)}&=&\frac{4\pi^{(d-1)/2}}{d\Gamma\left(\frac{d}{2}\right)}\sum_{j=1}^2\; n_j \mu_{ji}\sigma_{ij}^{d-1}\chi_{ij}
\Bigg(\frac{2T_i^{(0)}}{m_i}+\frac{2T_j^{(0)}}{m_j}\Bigg)^{1/2}\nonumber\\
& & \times (1+\al_{ij})\Bigg[1-\frac{\mu_{ji}}{2}(1+\al_{ij})\Big(1+\frac{m_i T_j^{(0)}}{m_j T_i^{(0)}}\Big)\Bigg].
\eeqa
It must be remarked that the theoretical results for the temperature ratio obtained by using the Maxwellian approximation \eqref{3.7} for the partial cooling rates shows in general an excellent agreement with Monte Carlo simulations. \cite{MG02,BT02,ChGG22}

\subsection{First-order approximation. Shear viscosity coefficient}

The analysis to first-order in the shear rate is large and tedious. As said before, as expected the GDH-theory yields a nonzero kinetic contribution $\eta_\text{k}$ to $\eta$, even for dilute systems. \cite{GD02,GMD06} The expression of $\eta$ is
\beq
\label{3.13}
\eta^{\text{GDH}}=\eta_\text{k}^{\text{GDH}}+\eta_\text{c}^{\text{GDH}},
\eeq
where
\beq
\label{3.14}
\eta_\text{k}^{\text{GDH}}=\sum_{i=1}^2\eta_i^{\text{k}}, \quad \eta_i^{\text{k}}=-\frac{m_i}{a}\int d\mathbf{V}V_xV_yf_i^{(1)}(\mathbf{V}),
\eeq
and \cite{G22}
\beqa
\label{3.15}
\eta_c^{\text{GDH}}&=&\frac{2\pi^{d/2}}{d(d+2)\Gamma\left(\frac{d}{2}\right)}\sum_{i=1}^2 \sum_{j=1}^2 n_i \sigma_{ij}^{d}\chi_{ij} \mu_{ij} (1+\al_{ij})\Bigg[\eta_j^k\nonumber\\
& &+n_j m_j \sigma_{ij}\Big(\frac{m_iT_j^{(0)}+m_jT_i^{(0)}}{2\pi m_i m_j}\Big)^{1/2}\Bigg].
\eeqa
As in the case of $\zeta_i^{(0)}$, upon obtaining Eq.\ \eqref{3.15} we have approximated $f_i^{(0)}(\mathbf{V})$ by the Maxwellian distribution $f_{i,\text{M}}(\mathbf{V})$. So far, the expression \eqref{3.14} for $\eta_i^{\text{k}}$ is exact. However, the distributions $f_1^{(0)}$ and $f_2^{(0)}$ obey a set of coupled linear integral equations which exact solution is not known to date. Therefore, as usual in molecular mixtures, \cite{CC70} we take the low order truncation of the series expansion of those distributions in Sonine polynomials. The leading Sonine approximation to $f_i^{(1)}(\mathbf{V})$ is
\begin{equation}
\label{3.16}
f_i^{(1)}(\mathbf{V}) \to -a\frac{m_i\eta_i^{\text{k}}}{n_iT_i^{(0)2}}V_xV_y f_{i,\text{M}}(\mathbf{V}).
\end{equation}
The kinetic coefficients $\eta_i^{\text{k}}$ can be computed from the Enskog kinetic equation for $f_i^{(1)}(\mathbf{V})$ by using the approximation \eqref{3.16}. After some algebra, one gets the expressions
\beq
\label{3.17}
\eta_1^{\text{k}}=\frac{\left(\tau_{22}-\zeta^{(0)}\right)A_1-\tau_{12} A_2}{\zeta^{(0)2}-\zeta^{(0)}\left(\tau_{11}+\tau_{22}\right)+\tau_{11}\tau_{22}-\tau_{12}\tau_{21}},
\eeq
\vspace{0.001cm}
\beq
\label{3.18}
\eta_2^{\text{k}}=\frac{\left(\tau_{11}-\zeta^{(0)}\right)A_2-\tau_{21} A_1}{\zeta^{(0)2}-\zeta^{(0)}\left(\tau_{11}+\tau_{22}\right)+\tau_{11}\tau_{22}-\tau_{12}\tau_{21}},
\eeq
where
\beqa
\label{3.19}
A_i&=&n_i T_i^{(0)}+\frac{\pi^{d/2}}{d(d+2)\Gamma\left(\frac{d}{2}\right)}\sum_{j=1}^2
n_in_j\sigma_{ij}^d m_{ij}\chi_{ij}(1+\al_{ij})\nonumber\\
& & \times \Bigg[\mu_{ji}\left(3\al_{ij}-1\right)
\Bigg(\frac{T_i^{(0)}}{m_i}+\frac{T_j^{(0)}}{m_j}\Bigg)-
4\frac{T_i^{(0)}-T_j^{(0)}}{m_i+m_j}\Bigg],\nonumber\\
\eeqa
\begin{widetext}
\beqa
\label{3.20}
\tau_{11}&=&\frac{2\pi^{(d-1)/2}}{d(d+2)\Gamma\left(\frac{d}{2}\right)}v_{\text{th}}\Bigg\{
n_1\sigma_{1}^{d-1}\chi_{11}(2\theta_1)^{-1/2}(3+2d-3\alpha_{11})(1+\alpha_{11}) +2n_2 \chi_{12}\mu_{21}(1+\alpha_{12})\theta_1^{3/2}\theta_2^{-1/2}\nonumber\\
& &\times \Big[
(d+3)(\mu_{12}\theta_2-\mu_{21}\theta_1)\theta_1^{-2}(\theta_1+\theta_2)^{-1/2}+\frac{3+2d-3\alpha_{12}}{2}\mu_{21}
\theta_1^{-2}(\theta_1+\theta_2)^{1/2}\nonumber\\
& &
+\frac{2d(d+1)-4}{2(d-1)}\theta_1^{-1}(\theta_1+\theta_2)^{-1/2}\Big]\Bigg\},
\eeqa
\beqa
\label{3.21}
\tau_{12}&=&\frac{4\pi^{(d-1)/2}}{d(d+2)\Gamma\left(\frac{d}{2}\right)}v_{\text{th}}
n_1\sigma_{12}^{d-1}\chi_{12}\mu_{12}\theta_1^{-1/2}\theta_2^{3/2}
(1+\alpha_{12})\Big[
(d+3)(\mu_{12}\theta_2-\mu_{21}\theta_1)\theta_2^{-2}(\theta_1+\theta_2)^{-1/2}
\nonumber\\
& &+\frac{3+2d-3\alpha_{12}}{2}\mu_{21}\theta_2^{-2}(\theta_1+\theta_2)^{1/2}
-\frac{2d(d+1)-4}{2(d-1)}\theta_2^{-1}(\theta_1+\theta_2)^{-1/2}\Big].
\eeqa
\end{widetext}
Here, $v_{\text{th}}=\sqrt{2T/m_{12}}$ is a thermal speed, $\theta_1=T/(\mu_{21}T_1^{(0)})$, and $\theta_2=T/(\mu_{12}T_2^{(0)})$. The forms of $\tau_{22}$ and $\tau_{21}$ can be easily obtained from Eqs.\ \eqref{3.20} and \eqref{3.21} by interchanging $1\leftrightarrow 2$.

For elastic collisions ($\al_{ij}=1$) and hard spheres ($d=3$), the expression of $\eta^{\text{GDH}}$ given by Eqs.\ \eqref{3.13}--\eqref{3.15} agree with the results derived many years ago from the Enskog kinetic theory for molecular mixtures of hard spheres. \cite{KS79b} However, the expression \eqref{3.3} of $\eta_c^{\text{SM}}$ provided by the SM-theory is inconsistent with the results for molecular mixtures. \cite{KS79b} Regarding the comparison between the SM-theory and GDH-theory for the collisional coefficient $\eta_c$, we see that both theories lead to different expressions even for elastic collisions. In this limit case ($\al_{ij}=1$), Eqs.\ \eqref{3.3} and \eqref{3.15} are equivalent only when the kinetic coefficients $\eta_i^\text{k}$ are neglected in the GDH-theory.

\section{Monte Carlo simulation of a granular binary mixture under SSF}
\label{sec5}

To assess the degree of accuracy of the SM and GDH theories one has to resort to computer simulations. More specifically, in this paper we have numerically solved the Enskog equation by means of the extension of the well-known DSMC method \cite{B94} to dense gases. The method is usually referred to as the ESMC method. \cite{MS96,MS97} In the simulations carried out in this paper the method has been slightly modified to determine the shear viscosity coefficient of a \emph{granular} binary mixture for moderate densities. One important advantage of using the ESMC method instead of molecular dynamics simulations is that the simulation method is easy to implement from a computational point of view due to the fact that the SSF state becomes spatially homogeneous in the local Lagrangian frame defined by the position $\mathbf{R}$ and the peculiar velocity $\mathbf{V}$.

As said in section \ref{sec1}, in the absence of a thermostating force ($\xi=0$), a granular fluid in the SSF reaches a steady state where the viscous heating effect is exactly compensated for by the collisional cooling. In this case, the SSF is inherently a non-Newtonian state. \cite{SGD04} Thus, to allow that the granular temperature grows in time due to the viscous heating effect (as in the case of elastic collisions), we excite the granular mixture by means of the Gaussian force
\beq
\label{5.0}
\mathbf{F}_i^{\text{th}}=\frac{1}{2}m_i \xi \mathbf{V}.
\eeq
According to Eq.\ \eqref{2.5}, if $\xi=\zeta$ then the Gaussian force exactly balances the energy lost by collisions. In this situation, since the collision frequency $\nu(t)$ for hard spheres is proportional to $\sqrt{T(t)}$, then the relevant uniformity parameter $a^*=a/\nu(t)$ (\emph{reduced} shear rate) monotonically decreases in time and so, the mixture asymptotically reaches a Navier--Stokes regime where the \emph{reduced} shear viscosity
\beq
\label{5.1}
\eta^*=-\lim_{t\to \infty}\frac{P_{xy}^*}{a^*}
\eeq
can be measured in the simulations. Here, $P_{xy}^*=P_{xy}/nT$,
\beq
\label{5.1.1}
\eta^*=\frac{\nu}{n T}\eta,
\eeq
and
\beq
\label{5.1.2}
\nu(T(t))=\sqrt{\pi}n \sigma_{12}^{d-1}v_\text{th}(T(t))
\eeq
is an effective collision frequency for hard spheres. In the case of molecular mixtures (where $\zeta=\xi=0$), Eq.\ \eqref{5.1} was employed by Naitoh and Ono \cite{NO79} to measure the Navier--Stokes shear viscosity $\eta$ of a hard-sphere gas. The same procedure can be followed for granular mixtures when the system is heated by the Gaussian thermostat. In this case, $\eta$ has been also measured in heated granular mixtures of low \cite{MG03} and moderate \cite{GM03} densities. Here, since we are mainly interested in assessing the SM and GDH theories at the level of the collisional coefficient $\eta_c$, our simulations will consider moderately dense mixtures where the Enskog equation applies.

The application of the ESMC method to the SFF state was made years ago by Montanero and Santos. \cite{MS96,MS97} It will be briefly presented herein for the physical case $d=3$; the interested reader is referred to Refs.\ \onlinecite{MG03,GM03} for a more complete description on the application of this simulation method to sheared granular mixtures. As usual in the ESMC method, the velocity distribution function of the species $i$ is represented by the peculiar velocities $\{{\bf V}_k\}$ of $N_i$ ``simulated" particles:
\begin{equation}
\label{5.2}
f_i({\bf V},t)\to n_i \frac{1}{N_i}\sum_{k=1}^{N_i} \delta({\bf V}-{\bf
V}_k(t))\; .
\end{equation}
Although the number of particles $N_i$ of the species $i$ is arbitrary, the relation $N_1/N_2=n_1/n_2$ must be considered. For the sake of simplicity,  one assigns initially velocities to the particles drawn from the Maxwell-Boltzmann probability distribution:
\begin{equation}
\label{5.3}
f_{i\text{M}}(\mathbf{V},0)=n_i\ \pi^{-3/2}\ V_{0i}^{-3}(0)\ \exp\left[-V^2/V_{0i}^2(0)\right]\;,
\end{equation}
where $V_{0i}^2(0)=2T(0)/m_i$ and $T(0)$ is the initial temperature. To enforce a vanishing initial total momentum, the velocity of every particle is subsequently subtracted by the amount $N_i^{-1} \sum_k {\bf V}_k(0)$.

The free motion and the collisions are uncoupled over a time step $\Delta t$; this time is small compared with both the mean free time and the inverse shear rate. Since the reduced shear rate $a^*$ decreases monotonically in time, the value of $\Delta t$ must be updated in the course of the simulation. Since the SSF state is \emph{homogeneous} in the local Lagrangian frame moving with the peculiar velocity $\mathbf{V}$, particles of each species ($i=1,2$) are subjected to the action of a non-conservative inertial force
\beq
\label{5.4}
F_{i,\lambda}=-m_i\ a_{\lambda\beta} V_\beta.
\eeq
Consequently, the free motion stage consists of making the change
\beq
\label{5.4.0}
\mathbf{V}_k\to \mathbf{V}_k-\mathsf{a}\cdot\mathbf{V}_k \Delta t.
\eeq

In the collision stage, binary interactions between particles of species $i$ and $j$ must be considered. Then, a sample of
\beq
\label{5.5}
\frac{1}{2} N_i \omega_{\text{max}}^{(ij)}\Delta t
\eeq
pairs is chosen at random with equiprobability to simulate the collisions between particles of species $i$ with $j$. In Eq.\ \eqref{5.5},  $\omega_{\text{max}}^{(ij)}$ is an upper bound estimate of the probability that a particle of the species $i$ collides with a particle of the species $j$. Let us consider a pair $(k,\ell)$ belonging to this sample ($k$ denotes a particle of species $i$ and $\ell$ a particle of species $j$). For each pair $(k,\ell)$ with velocities $(\mathbf{V}_k,\mathbf{V}_{\ell})$, the following steps are taken:
\begin{enumerate}
\item A given direction $\widehat{\boldsymbol{\sigma}}_{k\ell}$ is chosen at random with equiprobability.

\item The collision between particles $k$ and $\ell$ is accepted with a probability equal to $\Theta(\mathbf{g}_{k\ell}\cdot
\widehat{\boldsymbol{\sigma}}_{k\ell})\omega_{k\ell}^{(ij)}/ \omega_{\text{max}}^{(ij)}$, where $\omega_{k\ell}^{(ij)}=4\pi \sigma_{ij}^2 n_j
|\mathbf{g}_{k\ell}\cdot \widehat{\boldsymbol{\sigma}}_{k\ell}|$ and $\mathbf{g}_{k\ell}=\mathbf{V}_k-\mathbf{V}_{\ell}-\sigma_{ij} \mathsf{a}\cdot \widehat{\boldsymbol{\sigma}}_{k\ell}$ is the relative velocity of the colliding pair in the Lagrangian frame.

\item In the case that the collision is accepted, postcollisional velocities are assigned to both particles according to the scattering rules \eqref{2.3.1} and \eqref{2.3.2}:
\begin{equation}
\label{5.6}
\mathbf{V}_{k}\to \mathbf{V}_{k}-\mu_{ji}(1+\alpha_{ij})(\mathbf{g}_{k\ell}\cdot \widehat{\boldsymbol{\sigma}}_{k\ell})\widehat{\boldsymbol{\sigma}}_{k\ell},
\end{equation}
\begin{equation}
\label{5.7}
\mathbf{V}_{\ell}\to \mathbf{V}_{\ell}+\mu_{ij}(1+\alpha_{ij})(\mathbf{g}_{k\ell}\cdot \widehat{\boldsymbol{\sigma}}_{k\ell})\widehat{\boldsymbol{\sigma}}_{k\ell}.
\end{equation}
If in a collision $\omega_{k\ell}^{(ij)}>\omega_{\text{max}}^{(ij)}$, the estimate of $\omega_{\text{max}}^{(ij)}$ is updated as $\omega_{\text{max}}^{(ij)}=\omega_{k\ell}^{(ij)}$.
\end{enumerate}

The procedure described above is performed for $i=1,2$ and $j=1,2$. The granular temperature is calculated before and after the collision stage, and thus the instantaneous value of the cooling rate $\zeta$ is obtained. After the collisions have been calculated, the thermostat Gaussian force \eqref{5.0} is considered by making the change:
\beq
\label{5.7.1}
\mathbf{V}_k\to \mathbf{V}_k+\frac{1}{2} \zeta \mathbf{V}_k\Delta t.
\eeq

The kinetic and collisional transfer contributions to the pressure tensor are evaluated along the course of the simulations. They are given as
\begin{equation}
\label{5.8}
\mathsf{P}^{\text{k}}=\sum_{i=1}^{2} \frac{m_i n_i}{N_i}\sum_{k=1}^{N_i} \mathbf{V}_k \mathbf{V}_k,
\end{equation}
\begin{equation}
\label{5.9}
\mathsf{P}^{\text{c}}=\frac{n}{2N\Delta t}{\sum_{k\ell}}^{\dagger} \mu_{ij}m_j \sigma_{ij}(1+\alpha_{ij})
(\mathbf{g}_{k\ell}\cdot \widehat{\boldsymbol{\sigma}}_{k\ell})\widehat{\boldsymbol{\sigma}}_{k\ell}
\widehat{\boldsymbol{\sigma}}_{k\ell},
\end{equation}
where the dagger means that the summation is restricted to the accepted collisions \vicente{and subscripts $i$ and $j$ refer to the type of specie. Moreover, we recall that in Eqs.\ \eqref{5.8} and \eqref{5.9} the subscript $k$ refers to a particle of species $i$ while the subscript $\ell$ refers to a particle of species $j$}.

\vicente{As mentioned before, in our ESMC simulations, the velocities of the particles are  changed in each time step due to two uncoupled mechanisms: the free streaming stage [where all particle velocities are updated due to the shear rate and the Gaussian thermostat force following Eqs.\ \eqref{5.4.0} and \eqref{5.7.1}, respectively] and the collision stage [where only a selected sample of particles changes its velocities following Eqs.\ \eqref{5.6} and \eqref{5.7}]. This allows us to estimate separately the collisional and kinetic contributions to the pressure tensor in each time step. The former is obtained by summing only the contributions given by Eq.\ \eqref{5.9} of the selected collision pairs at the end of the collision stage once all collisions were performed. The kinetic contribution is computed by taking into account all the velocities of the particles independently if they are collided at the end of each time step once free streaming was applied to all particles. The kinetic and collisional contributions to the pressure tensor are averaged over an specific number of replicas $\mathcal{N}$. The pressure tensor is obtained from Eq.\ \eqref{2.8} while} the (reduced) shear viscosity $\eta$ is obtained from Eq.\ \eqref{5.1.1}.

One of the most determining steps for DSMC calculations is the use of an efficient pseudo-random numbers generator (PRNG). It is well-known that PRNGs create a long but finite sequence of pseudo-random numbers. The period of the sequence may play a major role in the quality of this generator as the amount of required random numbers grows up. In our simulations, use has been made of a larger number of particles, time steps and replicas than in the simulations carried out in previous works. \cite{MG03,GM03} For this reason, we have implemented the Mersienne Twister algorithm (MT19937), which is based on Mersienne prime numbers, developed by Matsumoto and Nishimura. \cite{MN98} The algorithm provides a set of uniform distributed pseudo-random numbers with an extremely massive period of $2^{19937}-1$ and 623-dimensional equidistribution up to 32-bit accuracy, while using a working area of only 624 words. In addition, the initial Maxwellian distributions were generated using the Marsaglia polar method \cite{MB64} for the Box-Muller transform. \cite{BM58}

Moreover, to improve the statistics, as said before the results have been averaged over a number $\mathcal{N}$ of independent realizations or replicas. PRNG was iniciated in each replica with a different seed in order to ensure the use of different pseudo-random number sequences. In our simulations we have typically taken a total number of particles $N=N_1+N_2=5\times 10^5$, a number of replicas $\mathcal{N}=20$, and a time step $\Delta t=5\times 10^{-3} \lambda_{11}/V_{01}(0)$. Here, $\lambda_{11}=(\sqrt{2} \pi n_1 \sigma_{11}^2)^{-1}$ is the mean free path for collisions 1--1 when $d=3$.

\begin{figure}
{\includegraphics[width=0.9\columnwidth]{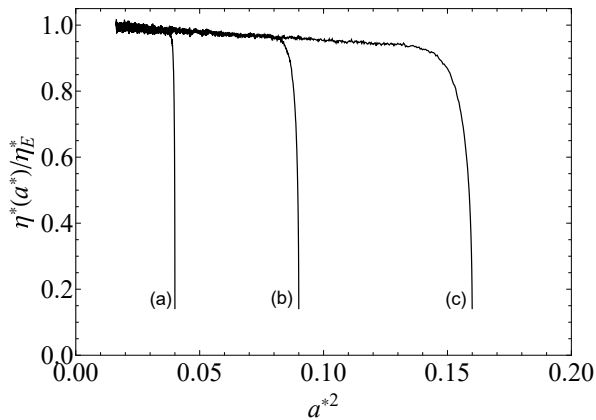}}
\caption{Plot of the ratio $\eta^*(a^*)/\eta_\text{E}^*$ versus $a^{*2}$ for a three-dimensional mixture with $x_1=\frac{1}{2}$, $\sigma_1/\sigma_2=2$, $m_1/m_2=4$, $\phi=0.1$, and a common coefficient of restitution $\al_{11}=\al_{22}=\al_{12}=0.8$. Three different values of the initial shear rate $a_0^*$ are considered: $a_0^*$=0.2 (a), $a_0^*$=0.3 (b), and $a_0^*$=0.4 (c). Here, $\eta_\text{E}^*$ refers to the value of the Navier--Stokes shear viscosity provided by the GDH-theory by solving the Enskog equation in the heated SSF state.
\label{fig1}}
\end{figure}

Before studying the dependence of $\eta^*$ on the parameter space of the mixture, it is convenient to gauge the reliability of the simulation method. In other words, for given values of the mass and diameter ratios, the coefficients of restitution, the concentration, and the density, the (reduced) shear viscosity $\eta^*$ must achieve a value independent of the initial conditions for long times (which is equivalent to the limit $a^*\to 0$). To illustrate this behavior, Fig.\ \ref{fig1} plots the ratio $\eta^*(a^*)/\eta_\text{E}^*$ for three different choices of the initial shear rate $a_0^*=a/\nu(T(0))$: 0.2, 0.3, and 0.4. Here, $\eta^*(a^*)$ refers to the value of the (reduced) shear viscosity measured in the simulations while $\eta_\text{E}^*$ corresponds to the theoretical Navier--Stokes value predicted by the GDH-theory. Here, we consider a three-dimensional mixture with $x_1=\frac{1}{2}$, $\sigma_1/\sigma_2=2$, $m_1/m_2=4$, $\phi=0.1$, and a common coefficient of restitution $\al_{11}=\al_{22}=\al_{12}=0.8$. Figure \ref{fig1} highlights the collapse of the three curves (corresponding to a three different initial conditions) to a common value after a transient period of a few mean free times. Consequently, a hydrodynamic regime independent of the initial preparation of the system is reached for sufficiently long times. As a byproduct, we also observe that there is a time window (which corresponds to the region of very small values of $a^{*2}$) where the ratio $\eta^*(a^*)/\eta_\text{E}^*$ fluctuates around 1. This means that the shear viscosity coefficient measured in the simulations when the (reduced) shear rate is small agrees very well with the one obtained from the Enskog equation by the GHD-theory. Similar behaviors have been found for other different mixtures. As remarked in Ref.\ \onlinecite{GM03}, note that the strict limit $a^*\to 0$ is not attainable in the simulations since it requires an infinite amount of time.

\section{Comparison between kinetic theories and computer simulations}
\label{sec6}

\begin{figure}
{\includegraphics[width=0.7\columnwidth]{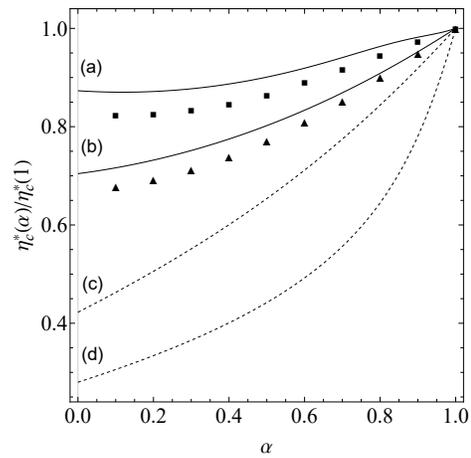}}
\caption{Dependence of $\eta_c^*(\al)/\eta_c^*(1)$ on the (common) coefficient of restitution $\al$ for $d=3$, $x_1=\frac{1}{2}$, $\sigma_1/\sigma_2=\frac{1}{2}$, $\phi=0.2$, and two different values of the mass ratio: $m_1/m_2=10$ [lines (a) and (c) and squares] and $m_1/m_2=2$ [lines (b) and (d) and triangles]. The solid lines  correspond to the GDH-theory whereas the dashed lines refer to the SM-theory. Here, $\eta_c^*(1)$ corresponds to the (dimensionless) collisional contribution to $\eta^*$ for elastic collisions.
\label{fig2}}
\end{figure}

Once the consistency of the simulation method to measure the shear viscosity in a heated granular mixture has been tested, we want to analyze the dependence of $\eta$ on the parameters of the mixture. More specifically, since the SM-theory is focussed essentially in the collisional contribution $\eta_c$ to $\eta$, we compare in this section the predictions of both kinetic theories (the SM- and GDH-theories) for $\eta_c$ with the results obtained from the ESMC method. On the other hand, since a complete presentation of the results is complex due to the high number of parameters involved in the problem, henceforth we will assume a three-dimensional ($d=3$) mixture for $\eta_c$ with a concentration of $x_1=\frac{1}{2}$ and constituted by spheres made of the same material $\al_{11}=\al_{22}=\al_{12}\equiv \al$. This reduces the number of parameters to four quantities: $\left\{\sigma_1/\sigma_2, m_1/m_2, \phi, \al \right\}$. In the case of hard-spheres ($d=3$), a good approximation for the pair correlation functions $\chi_{ij}$ is \cite{GH72}
\beq
\label{6.1}
\chi_{ij}=\frac{1}{1-\phi}+\frac{3}{2}\frac{\phi}{(1-\phi)^2}\frac{\sigma_i\sigma_jM_2}{\sigma_{ij}M_3}+
\frac{1}{2}\frac{\phi^2}{(1-\phi)^3}\Bigg(\frac{\sigma_i\sigma_jM_2}{\sigma_{ij}M_3}\Bigg)^2,
\eeq
where $M_s=\sum_i x_i \sigma_i^s$.

\begin{figure}
{\includegraphics[width=0.7\columnwidth]{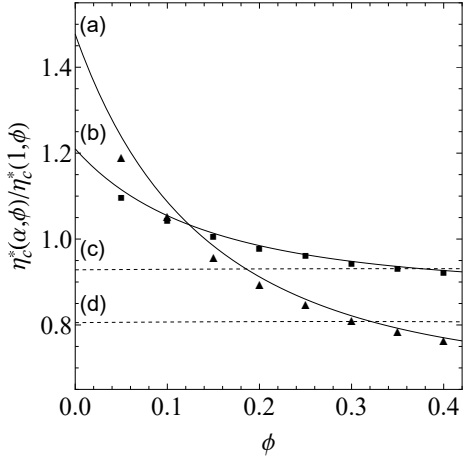}}
\caption{Plot of $\eta_c^*(\al,\phi)/\eta_c^*(1,\phi)$ versus the volume fraction $\phi$ for $d=3$, $x_1=\frac{1}{2}$, $\sigma_1/\sigma_2=2$, $m_1/m_2=10$, and two values of $\al$: $\al=0.5$ [the solid line (a) is for the GDH-theory and the dashed line (c) is for the SM-theory] and $\al=0.8$ [the solid line (b) is for the GDH-theory and the dashed line (d) is for the SM-theory]. The symbols correspond to the ESMC results: triangles for $\al=0.5$ and squares for $\al=0.8$.
\label{fig3}}
\end{figure}

Figure \ref{fig2} shows the dependence of the ratio $\eta_c^*(\al)/\eta_c^*(1)$ on the coefficient of restitution $\al$ for a binary mixture with $\sigma_1/\sigma_2=\frac{1}{2}$, $\phi=0.2$, and two different values of the mass ratio $m_1/m_2$. Here, $\eta_c^*(1)$ refers to the value of the collisional shear vicosity for elastic collisions. Although the SM-theory reproduces qualitatively well the $\al$-dependence of $\eta_c^*$ (this coefficient decreases with increasing inelasticity), significant quantitative discrepancies with simulations appear specially for strong dissipation. On the other hand, comparison between the GDH-theory and computer simulations shows a much better agreement than the one found for the SM-theory. \vicente{In fact, the results obtained from the GDH-theory (which we recall have been derived by considering the first-Sonine approximation) compares in general very well with simulation data. The differences between this theory and ESMC results tend to increase slightly as the coefficient of restitution decreases. We also observe that the first-Sonine solution to $\eta^*$ overestimates the results obtained from computer simulations. In this context and based on previous results derived for the tracer diffusion coefficient, \cite{GM04} [see Figs.\ 6.4 and 6.5 of Ref.\ \onlinecite{G19}] we expect that the discrepancies between the first-Sonine approximation and simulations can be in part mitigated by considering the second-Sonine approximation to $\eta^*$. We plan to perform this quite long and tedious calculation in the near future.}

More significant discrepancies between the SM and GDH kinetic theories appear when one considers the dependence of the ratio $\eta_c^*(\al,\phi)/\eta_c^*(1,\phi)$ on the total solid volume fraction $\phi$. This is shown in Fig.\ \ref{fig3} where $\eta_c^*(\al,\phi)/\eta_c^*(1,\phi)$ is plotted versus $\phi$ for a mixture with $\sigma_1/\sigma_2=2$, $m_1/m_2=10$, and two values of $\al$. As already remarked in Ref.\ \onlinecite{G21}, while the SM-theory shows a very weak density dependence of the above ratio for any value of $\al$, the GDH-theory clearly shows a significant decreasing of $\eta_c^*(\al,\phi)/\eta_c^*(1,\phi)$ as density increases, regardless of the value of the coefficient of restitution. With respect to the comparison with Monte Carlo simulations, we observe an excellent agreement between the theoretical predictions of the GDH-theory and the simulation data over the entire range of values of the solid volume fraction considered.

The tiny dependence of the ratio $\eta_c^{*\text{SM}}(\al,\phi)/\eta_c^{*\text{SM}}(1,\phi)$ on $\phi$ at a given value of $\al$ in the SM-theory can be explained by the fact that the only dependence of this ratio on $\phi$ in this theory is via the partial temperatures $T_i^{(0)}$, whose dependence on $\phi$ is very small. However, the dependence of $\eta_c^{*\text{GDH}}(\al,\phi)/\eta_c^{*\text{GDH}}(1,\phi)$ on $\phi$ in the GDH-theory is not only through $T_i^{(0)}$ but also through the kinetic coefficients $\eta_i^{\text{k}}$. To show it in a more clean way, it is quite instructive to consider the limiting case of mechanically equivalent particles ($m_1=m_2$, $\sigma_1=\sigma_2$, and $\al_{ij}=\al$). In this limit case,
\beq
\label{6.2}
\frac{\eta_c^{*\text{SM}}(\al,\phi)}{\eta_c^{*\text{SM}}(1,\phi)}=\frac{1+\al}{2},
\eeq
while
\beq
\label{6.3}
\frac{\eta_c^{*\text{GDH}}(\al,\phi)}{\eta_c^{*\text{GDH}}(1,\phi)}=\frac{1+\al}{2} A(\phi,\al),
\eeq
where the function $A(\phi,\al)$ has a complex dependence on both $\phi$ and $\al$. For the sake of illustration, for $d=3$, $A(\phi,\al)$ is given by
\beq
\label{6.4}
A(\phi,\al)=\frac{1+B(\phi,\al)}{1+C(\phi)},
\eeq
where
\beq
\label{6.5}
B(\phi,\al)=\frac{5\pi}{16 \phi \chi}\frac{1-\frac{2}{5}(1+\al)(1-3\al)\phi \chi}{(1+\al)(2+\al)},
\eeq
\beq
\label{6.6}
C(\phi)=\frac{5\pi}{96 \phi \chi}\left(1+\frac{8}{5}\phi \chi\right).
\eeq

\begin{figure}
{\includegraphics[width=0.7\columnwidth]{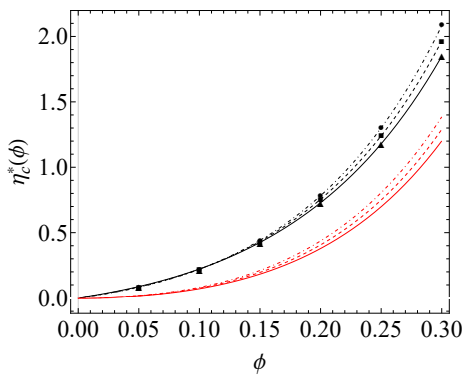}}
\caption{Plot of the (dimensionless) collisional shear viscosity $\eta_c^*$ as a function of the solid volume fraction $\phi$ for $m_1/m_2=4$, $\sigma_1/\sigma_2=1$, and three different values of the coefficient of restitution $\al$: $\al=0.9$ (solid lines and triangles), $\al=0.8$ (dashed lines and squares), and $\al=0.7$ (dotted lines and circles). The black lines correspond to the GDH-theory while the red lines are for the SM-theory. The symbols refer to the results obtained from the ESMC method.
\label{fig4}}
\end{figure}

\begin{figure}
{\includegraphics[width=0.7\columnwidth]{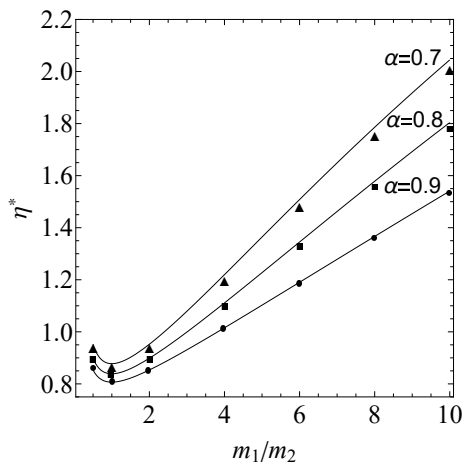}}
\caption{Plot of the (dimensionless) shear viscosity $\eta^*=\eta_k^*+\eta_c^*$ as a function of the mass ratio $m_1/m_2$, for $\sigma_1/\sigma_2=1$, $\phi=0.1$, and three different values of $\al$. The linea are the theoretical predictions of the GDH-thoery adn the symbols correspond to the results obtained from the ESMC method.
\label{fig5}}
\end{figure}

To complement Fig.\ \ref{fig3}, Fig.\ \ref{fig4} shows the $\phi$-dependence of the (dimensionless) collisional shear viscosity $\eta_c^*$ for $m_1/m_2=4$, $\sigma_1/\sigma_2=1$, and three different values of the coefficient of restitution $\al$. As for Fig.\ \ref{fig3}, the good agreement between the GDH-theory and Monte Carlo simulations indicates again that the collisional transfer contributions to the shear viscosity are provided accurately by Eq.\ \eqref{3.15}. Important discrepancies between the expression \eqref{3.3} of the SM-theory and computer simulations are again observed, specially for high densities.

Although the main goal of the present paper is to asses the reliability of the SM-theory and GDH-theory at the level of the collisional coefficient $\eta_c^*$, it is also interesting to gauge the accuracy of the GDH-theory for the total shear viscosity $\eta^*=\eta_k^*+\eta_c^*$. Given that the SM-theory predicts $\eta_k^*=0$, we have not considered appropriate to include the SM-theory in this comparison. Figure \ref{fig5} shows $\eta^*$ versus the mass ratio $m_1/m_2$ for $\sigma_1/\sigma_2=1$, $\phi=0.1$, and three different values of $\al$. We observe first that the agreement between the GDH-theory and computer simulations is in general very good, although the differences between the theoretical and ESMC results tend to increase as inelasticity increases. As noted in Ref.\ \onlinecite{GM03}, at a given value of $\al$, we observe that $\eta^*$ exhibits a non-monotonic dependence on the mass ratio.

\section{Concluding remarks}
\label{sec7}

The main objective of this paper has been to assess the accuracy of two different kinetic theories for granular mixtures: the SM-theory \cite{SM21} and the GDH-theory. \cite{GDH07,GHD07} While the SM-theory is based on the assumption of Maxwellian distributions at different temperatures $T_i$ and velocities $\mathbf{U}_i$ for the true distribution functions $f_i(\mathbf{r}, \mathbf{v};t)$, the GDH-theory solves the Enskog kinetic equation by means of the application of the Chapman--Enskog method to first-order in spatial gradients. Due to the Maxwellian approximation of the SM-theory, it yields vanishing Navier--Stokes transport coefficients for dilute granular mixtures. \cite{GD02,GMD06} This is an important limitation of this theory. Thus, one expects that the SM-theory provides at least acceptable estimates for the collisional contributions to the transport coefficients.

A previous comparison \cite{G21} between both kinetic theories have shown important differences between them at the level of the collisional shear viscosity $\eta_c$, specially for strong inelasticity. To assess the reliability of each one of the theories, we have compared in this paper their theoretical predictions with those obtained by means of Monte Carlo simulations. More specifically, we have performed new simulations of moderately dense granular binary mixtures under SSF. As in previous works, \cite{MG03,GM03} we have introduced in the simulations an external thermostat force (proportional to the particle velocity) that supplies energy to the system to exactly compensate for the energy lost in collisions. In this way, the shearing work still heats the mixture so that, the reduced shear rate $a^*(t)=a/\nu(t)$ tends to zero in the long-time limit. Under these conditions, the system reaches a linear hydrodynamic regime where the Navier--Stokes shear viscosity of a heated granular binary mixture can be identified and measured in the simulations.

To reduce the number of independent parameters involved in the problem, the simulations have been carried for three-dimensional mixtures ($d=3$), with a mole fraction $x_1=\frac{1}{2}$ and with a (common) coefficient of normal restitution $\al\equiv \al_{rs}$. This reduces the number of relevant parameters to four ($\sigma_1/\sigma_2$, $m_1/m_2$, $\phi$, and $\al$). As expected, the comparison with computer simulations for $\eta_c$ have shown that the GDH-theory exhibits a much more better agreement with the ESMC results than the SM-theory. This is clearly shown in Fig.\ \ref{fig3} where the scaled coefficient $\eta_c^*(\al,\phi)/\eta_c^*(1,\phi)$ is plotted versus the density $\phi$ for two different values of $\al$. While the SM-theory predicts a tiny influence of $\phi$ on this coefficient, the GDH-theory shows that $\eta_c^*(\al,\phi)/\eta_c^*(1,\phi)$ decreases significantly with increasing density at a given value of the coefficient of restitution. On the other hand, in spite of the deficiencies of the SM-theory, it captures at least the $\al$-dependence of $\eta_c^*(\al,\phi)/\eta_c^*(1,\phi)$ for given values of density (see Fig.\ \ref{fig2}).

\vicente{As mentioned in Sec.\ \ref{sec6}, the differences between the first-Sonine approximation to $\eta^*$ and computer simulations could be in principle diminished by considering the second Sonine correction to the first-order distribution function. Although we do not have an evidence on the convergence of the second-Sonine approximation to the ESMC results in the SSF problem, previous works \cite{GM04} on the tracer diffusion coefficient seem to indicate that this approximation could mitigate the (small) discrepancies observed in this paper between the GDH-theory and simulations. Since the determination of the second-Sonine approximation to the shear viscosity involves a significant work, we expect to provide a support of the above assertion in a next work.}

\vicente{It is quite apparent that the accuracy of the SM and GDH theories have been assessed through a comparison with an ``exact'' numerical solution of the Enskog equation in the SSF obtained from the ESMC method.\cite{MS96,MS97} This method (which is an extension to dense gases of the well-known DSMC method\cite{B94}) is based on the same assumptions as the Enskog kinetic equation: (i) molecular motion and collisions are decoupled and (ii) absence of velocity correlations between the particles which are about to collide (molecular chaos hypothesis). A much more stringent assessment of the above kinetic theories could be made via a comparison with the results obtained from molecular dynamics simulations (which do not rely on any of the above assumptions). In this context, it is remarkable to note that the GDH-theory has been also tested with molecular dynamics simulations in a relatively complex problem: hydrodynamic instabilities in a transient, polydisperse granular system at moderate density with significant inelasticity levels.\cite{MGH14} The comparison between the theoretical predictions of the GDH-theory for the critical length scale $L_c$ (which expression involves the shear viscosity coefficient) and molecular dynamics results shows in general an excellent agreement in flows of strong dissipation ($\al_{ij}\geq 0.7$) and moderate solid volume fractions ($\phi\leq 0.2$). This good agreement between molecular dynamics and linear hydrodynamics (with the Navier--Stokes transport coefficients derived from the GDH-theory in the first-Sonine approximation) for the onset of velocity vortices must be considered as a nontrivial test of the reliability of kinetic theory for describing granular polydisperse flows even for strong inelasticity, finite density, and particle dissimilarity.}

\vicente{One of the main limitations of the present study is its restriction to the shear viscosity coefficient. This coefficient has been identified in computer simulations thanks to the simplicity of the SSF: a nonequilibrium state that becomes \emph{homogeneous} in the Lagrangian frame moving with the velocities of particles. This fact allows us to measure in a clean way the dependence of the Navier--Stokes shear viscosity coefficient on the parameter space of the system. As said before, the reliability of the GDH-theory has been also assessed in the computation of the critical length $L_c$ for the onset of instabilities in the homogeneous cooling state. \cite{MGH14} Needless to say, the assessment of other relevant transport coefficients of granular mixtures is still an open challenging issue. Among them, the thermal conductivity coefficient (whose collisional transfer contribution is different from zero at moderate densities) can be the next coefficient to be measured in computer simulations. However, its determination in the Navier--Stokes domain is a quite difficult problem due essentially to the coupling present in \emph{steady} states for granular gases between spatial gradients and collisional cooling.\cite{G19} In principle, two different strategies can be followed to get this coefficient. The first option would be the use of Green--Kubo relations. \cite{BR04,BRMG05} However, before carrying on simulations, one should first derive theoretically these relations for granular mixtures. As a second option and based on previous results obtained for \emph{dilute} monocomponent granular gases,\cite{MSG07} one could apply a  homogeneous, anisotropic velocity-dependent external force which produces heat flux in the absence of gradients. On the other hand, although this second option seems to be more reachable than the first one (since the transport coefficient is measured in homogeneous conditions), its fine tuning for \emph{dense} granular mixtures still requires a significant additional work which goes beyond the objective of the present paper. We plan to work on the last line in the near future}.

\acknowledgments

The authors acknowledge financial support from Grant PID2020-112936GB-I00 funded byMCIN/AEI/ 10.13039/501100011033, and from Grants IB20079 and GR18079 funded by Junta de Extremadura (Spain) and by ERDF A way of making Europe. All numerical calculations were performed on the Lusitania II computer of COMPUTAEX  funded by Junta de Extremadura (Spain).

\vspace{0.1cm}

\textbf{DATA AVAILABILITY}
The data that support the findings of this study are available from the corresponding author upon reasonable request.


%
\end{document}